\documentclass[useAMS,usenatbib]{mn2e}
\usepackage{times,amsmath,amssymb,txfonts,graphicx,aas_macros,natbib}
\usepackage{xspace,color}
\usepackage{ulem}

\renewcommand\emph[1]{\textit{#1}}

\usepackage{relsize}

\newcommand\Myr{\,\rm Myr}
\newcommand\Gyr{\,\rm Gyr}

\newcommand\kms{\,\rm km\,s^{-1}}
\newcommand\pc{\,\rm pc}

\newcommand\kpc{\,\rm kpc}

\newcommand\urms{u_{\rm rms}}

\newcommand\mB{\mn{\mathbf{B}}}
\newcommand\EMF{\mbox{\boldmath{${\cal E}$}}}

\newcommand{\mn}[1]{\overline{#1}}
\newcommand{\rms}[1]{\left<\right.\!#1\!\left.\right>}

\newcommand\Rm{\mathrm{Rm}}

\newcommand{\simgt}%
           {\,\hbox{\lower0.35ex\hbox{$\sim$}\llap{\raise0.35ex\hbox{$>$}}}\,}
\newcommand{\simlt}%
           {\,\hbox{\lower0.35ex\hbox{$\sim$}\llap{\raise0.35ex\hbox{$<$}}}\,}

\newcommand\NIRVANA{\textsc{Nirvana}\xspace}
\newcommand\NIII{\textsc{Nirvana-iii}\xspace}

\title[Quenching in ISM turbulence]%
      {On the magnetic quenching of mean-field effects 
        in supersonic interstellar turbulence}
      \author[Gressel, Bendre \& Elstner]%
             { Oliver~Gressel$^1$\thanks{E-mail:~oliver.gressel@nordita.org},
               Abhijit~Bendre$^2$ and
               Detlef~Elstner$^2$\\
               $^1$NORDITA, KTH Royal Institute of Technology and 
               Stockholm University,
               Roslagstullsbacken 23, 106 91 Stockholm, Sweden\\
               $^2$Leibniz-Institut f{\"u}r Astrophysik Potsdam (AIP), 
               An der Sternwarte 16, 14482, Potsdam, Germany}

\begin{document}

\date{Accepted 1988 December 15. %
      Received 1988 December 14; %
      in original form 1988 October 11}

\pagerange{\pageref{firstpage}--\pageref{lastpage}} \pubyear{2012}

\maketitle

\label{firstpage}

% ------------------------------------------------------------------------------

\begin{abstract}
%context
The emergence of large-scale magnetic fields observed in the diffuse
interstellar medium is explained by a turbulent dynamo. The underlying
transport coefficients have previously been extracted from numerical
simulations.
%aims
So far, this was restricted to the \emph{kinematic} regime, but we aim
to extend our analysis into the realm of dynamically important
fields. This marks an important step on which derived mean-field
models rely to explain observed equipartition-strength fields.
%methods
As in previous work, we diagnose turbulent transport coefficients by
means of the test-field method.
%results
We derive quenching functions for the dynamo $\alpha$~effect,
diamagnetic pumping, and turbulent diffusivity, which are compared
with theoretical expectations. At late times, we observe the
suppression of the vertical wind.
%conclusions
Because this potentially affects the removal of small-scale magnetic
helicity, new concerns arise about circumventing constraints imposed
by the conservation of magnetic helicity at high magnetic Reynolds
numbers. While present results cannot safely rule out this
possibility, the issue only becomes important at late stages and is
absent when the dynamo is quenched by the wind itself.
\end{abstract}


% ------------------------------------------------------------------------------

\section{Introduction}
\label{sec:intro}

The presence of Galactic rotation and vertical stratification renders
interstellar turbulence both inhomogeneous and anisotropic. It is well
known \citep[e.g.][]{1980mfmd.book.....K} that, in the framework of
Reynolds-averaged equations, this will lead to the emergence of
mean-field effects feeding on the spatial variations of the turbulence
intensity and density field. Different theoretical predictions
\citep{1993A&A...269..581R,1998A&A...335..488F} for the leading-order
effects in the \emph{kinematic} regime have previously been tested by
means of simulations \citep{2008AN....329..619G}. We now extend this
approach into the domain of dynamically important magnetic fields. The
main aim of the present work is to extract the intricately non-linear
dependence of the measured turbulence effects on the strength of the
Lorentz force -- this is usually referred to as ``magnetic
quenching''. The knowledge of this functional dependence is required
to evolve derived mean-field models into saturation. This is
important, because most observed spiral galaxies are found to harbour
large-scale magnetic fields of near-equipartition strength
\citep[see][for a topical compilation]{2010ASP..conf..197F}.
Mean-field models are required to predict global symmetries of the
magnetic field \citep{1996ARA&A..34..155B}, which is neither possible
in local resolved simulations, nor in global simulations ignoring
effects occurring on unresolved scales. There has been an active
debate \citep[see chapter~9 in][hereafter BS05]{2005PhR...417....1B}
about severe constraints inflicted upon the saturation level of the
mean-field dynamo, first addressed by \citet{1992ApJ...393..165V}.
These, potentially ``catastrophic'', constraints can arise as a result
of the conservation of magnetic helicity at large magnetic Reynolds
numbers, $\Rm$. However, \citet{2002A&A...387..453K} and later
\citet{2006A&A...448L..33S} have demonstrated that such constraints
can be alleviated by means of magnetic helicity \emph{fluxes}. While
such fluxes are likely present in our simulations in the form of
advection \citep[cf.][]{2008A&A...486L..35G}, this may be restricted
to the kinematic phase. For strong fields, we indeed observe the wind
advection to be quenched itself in certain cases, and this will
require careful attention. In the present paper, we will however
ignore this complication and focus on investigating the transition
into the quenched regime. We will furthermore try to distinguish
between different kinds of non-linearity that have been proposed.

% ------------------------------------------------------------------------------

\vspace{-1ex}
\section{Methods}
\label{sec:methods}

\begin{figure*}
  \center\includegraphics[width=1.8\columnwidth]{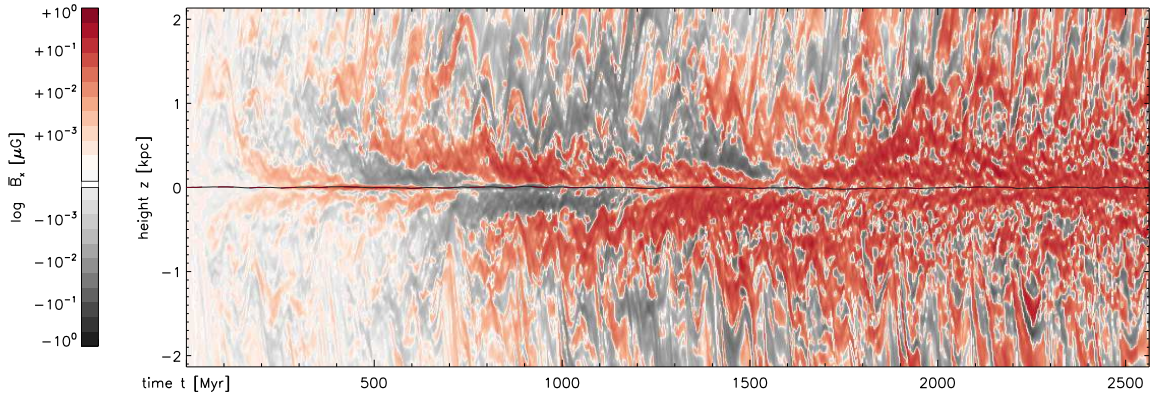}
  \caption{Space-time diagram of the radial magnetic field
    component. Colour coding shows the signed logarithm of the field
    strength. Note the evident ballistic trajectories, which are
    highly indicative of a Galactic fountain flow. Around $t=1.2\Gyr$,
    the prevalent vertical parity changes from dipolar to
    quadrupolar. }
  \label{fig:spctm}
\end{figure*}

We employ the \NIII MHD code \citep{2004JCoPh.196..393Z} to perform
direct simulations of the multi-phase interstellar medium (ISM) in a
local patch ($L_x=L_y=0.8\kpc$) of the Galaxy, but retaining the
vertical structure of the disc ($z\in\pm2.1\kpc$). The grid resolution
is $\Delta=8.3\pc$, and we combine shearing-periodic boundary
conditions in the plane \citep{2007CoPhC.176..652G} with outflow at
the vertical domain boundaries, where magnetic fields obey a
normal-field condition. By means of localised injection of thermal
energy, we mimic the effect of supernova explosions, thereby driving
strongly supersonic turbulence. Our numerical setup is similar to our
previous work but we here follow the evolution of the magnetic field
into its saturation. During the course of the simulation, we employ
the test-field (TF) method first introduced by
\citet{2005AN....326..245S} to infer turbulent closure parameters such
as the $\alpha$~effect, turbulent diamagnetism, and turbulent
diffusivity. The standard flavour of this method
\citep{2007GApFD.101...81S} is now usually referred to as
``quasi-kinematic'' TF method \citep{2010A&A...520A..28R}, as it has
been shown to remain fully valid into the non-kinematic regime in the
absence of magnetic background fluctuations
\citep{2008ApJ...687L..49B}. Since our driving is dominated by kinetic
forcing and we likely remain sub-critical to exciting a small-scale
dynamo \citep[see][]{2012ApJ...753...32M}, this is warranted in our
application. Whether, at higher $\Rm$, the presence of a fluctuation
dynamo will affect the evolution of the mean fields, remains open
\citep*{2012SSRv..tmp...57B}.

We measure TF coefficients as a function of time, $t$, and vertical
position, $z$. Due to the vertical stratification, the amplitude of
the $\alpha$~effect significantly depends on position, as does the
strength of the magnetic field. In other words, our system (unlike,
e.g., forced turbulence in an unstratified triply-periodic box) has a
rather poorly controlled background state and hence suffers from
strong random fluctuations. Because of this, it is not straightforward
to extract the quenching function. If we were to compile a naive
scatter plot (as is usually done in a homogeneous setting) we are
bound to confuse variations in $\alpha$ caused by fluctuations in the
background gradients (i.e. in $\nabla\rho$ and $\nabla \urms^2$) with
the attenuation due to the non-linear quenching. To resolve this
degeneracy, we have to make an assumption about the systematic
variation with height. A very simple possibility is to approximate the
vertical profiles by a ``power series'' around $z=0$, and for reasons
of simplicity, we restrict ourselves to the lowest-order variation in
$z$. For the $\alpha$~effect, and for the diamagnetic pumping,
$\gamma_z$, we accordingly measure linear slopes within a defined
central range in $z$. This is guided by inspection of
Fig.~\ref{fig:tf_coeff} presented in the following section. The
turbulent diffusivity, $\eta_{\rm T}$, of course, has a dominant even
parity with respect to $z=0$, and we accordingly apply a quadratic
fit.

The vertical profiles of the magnetic and kinetic energy densities are
found to be roughly bell-shaped. As a measure for the relative
importance of the magnetic field, we introduce the quantity
$\beta\equiv\rms{\bar{B}/B_{\rm eq}}$ as their ratio; here we have
used $\bar{B}\equiv |\mB| = (\bar{B}_R^2 + \bar{B}_\phi^2)^{1/2}$, and
$B_{\rm eq} \equiv (\mu_0\rho)^{1/2} \,\urms$, with $\mu_0$ the
permeability of free space, and where overbars denote horizontal
averages and angle brackets refer to rms~values when averaging over
$(z,t)$. We remark that the weak vertical field, $\bar{B}_z$, remains
constant throughout the run (a result of the periodic horizontal
boundaries, conserving vertical flux), and hence does not influence
the quenching function.

Due to strong fluctuations within the data, and to obtain reasonable
statistics, we bin the data into larger sections, for which we infer
data points $\alpha(\beta)$, $\gamma_z(\beta)$, and $\eta_{\rm
  T}(\beta)$ via linear regression as described above. Based on these
data, we then attempt to fit the commonly adopted algebraic quenching
formula \citep[see e.g.][]{1977SvA....21..479I}
\begin{equation}
  \alpha(\beta)=\alpha_0\,\frac{1}{1+q_\alpha\,\beta^{p_\alpha}}\,,
  \label{eq:quen}
\end{equation}
with 3 independent parameters $\alpha_0$, $q_\alpha$, and $p_\alpha$
in the case of $\alpha(\beta)$, and with analogous expressions for
$\gamma_z(\beta)$ and $\eta_{\rm T}(\beta)$, respectively. The
following meanings can be ascribed to the three coefficients each: The
subscript `0' refers to the unquenched amplitude, while the power $p$
determines the abruptness of the transition into the quenched regime;
$q$, taken to the power of $-1/p$, can be interpreted as the
(approximate) relative amplitude $\beta$ at which the quenching sets
in. The latter observation points at a potential correlation between
these two parameters when fitting the data, and this is indeed
confirmed for the case of $\alpha(\beta)$.

% ------------------------------------------------------------------------------

\vspace{-1ex}
\section{Results}
\label{sec:results}

The model presented in this work is identical to model ``Q4'' in
\citet{2008A&A...486L..35G}, with the exception that it is initialised
with a nano-Gauss \emph{vertical} field rather than a horizontal
one.\footnote{Strictly speaking, $\bar{B}_z$ is not a seed field but
  an imposed field. However, it induces rms~fluctuations in
  $\bar{B}_R$, and $\bar{B}_\phi$ which then act as seed.} The key
parameters of the simulation are a supernova rate of a quarter the
Galactic reference value of $\sigma_0=30\Myr^{-1}\kpc^2$, a rotation
rate of four times $25\kms\kpc^{-1}$, and background shear, $q=-1$,
equivalent to a flat rotation curve. For further details, we refer the
reader to the extensive description in \citet{2009PhDT........99G}. In
the present model, the inclusion of a net-vertical flux was chosen to
investigate a potential contribution from the magnetorotational
instability \citep[see][]{2005ApJ...629..849P,2010AN....331...34K}. We
here ignore this issue and focus on the quenching behaviour. The
effect of different field topologies will be discussed in a
forthcoming publication.

\subsection{Magnetic field evolution} % ---

For our purposes, the only noticeable effect of the vertical seed
field is that the growing mode takes a dipole-like symmetry with
respect to the midplane, $z=0$. This can be seen in
Fig.~\ref{fig:spctm}, where we show the spatiotemporal structure of
the horizontally averaged mean radial field. Ultimately, the
quadrupolar-like vertical symmetry will dominate, but not until the
parity of the field has reversed twice. Figure~\ref{fig:spctm} also
reveals a remarkable feature caused by the multi-phase nature of the
interstellar plasma: the occurrence of return flows, tracing-out the
Galactic fountain. Due to shock compression, the density is positively
correlated with the magnetic field strength. As the dense clumps rain
down into the Galactic potential, the frozen-in field gets dragged
along. A more comprehensive analysis of the correlation of mean
magnetic fields with the ISM phases will require more sophisticated
methods such as the ones recently proposed by
\citet{2012arXiv1206.6784G}.

\begin{figure}
  \center\includegraphics[width=0.92\columnwidth]{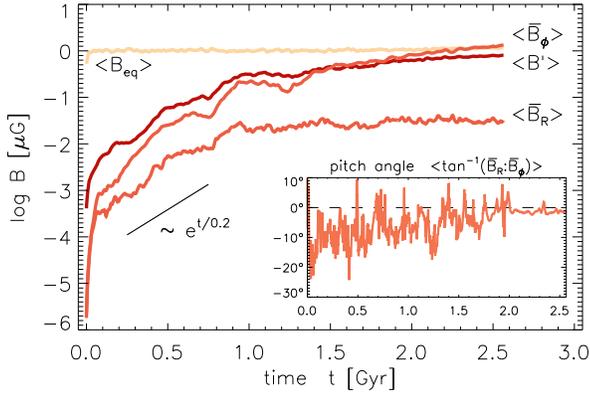}
  \caption{Time evolution of the fluctuating and regular magnetic
    field components, where overbars denote horizontal averages, and
    angle brackets refer to the rms variation within $\bar{B}_R(z)$,
    and $\bar{B}_\phi(z)$ at a given point in time. Note that the
    radial pitch angle (inset) is suppressed during saturation.}
  \label{fig:time_bfield}
\end{figure}

If we apply a further rms average over the vertical direction, we
obtain an estimate for the amplitude of the growing field, which is
shown in Fig.~\ref{fig:time_bfield} according to different
constituents. The turbulent velocity $\urms$ reaches a statistically
stationary state after less than $100\Myr$, and with it the
equipartition field $\rms{B_{\rm eq}}$. The fluctuating field
$\rms{B'}$, and the radial and azimuthal mean fields show an extended
phase of exponential growth, as indicated by the straight line. After
the saturation of the mean-field dynamo (evident from
$\rms{\bar{B}_R}$ remaining constant), we observe that
$\rms{\bar{B}_\phi}$ keeps growing due to stretching by the
differential rotation. As is shown in the inset of
Fig.~\ref{fig:time_bfield}, this has the consequence of a negligible
radial pitch angle in the saturated state -- which is at odds with
observational evidence of significant pitch angles for
equipartition-strength regular fields
\citep[see][]{2012SSRv..166..215B}. While this behaviour is
characteristic for dynamos of the $\alpha\Omega$~type, it has been
argued that the loss of pitch angle can be circumvented if the dynamo
is instead saturated by a wind \citep{2009IAUS..259..467E}. As we will
substantiate shortly, this is not the case for the model ``Q4''
discussed here, but has in fact been observed at higher supernova
rates (paper in preparation).

\subsection{Derivation of quenching functions} % ---

\begin{figure}
  \center\includegraphics[width=0.92\columnwidth]{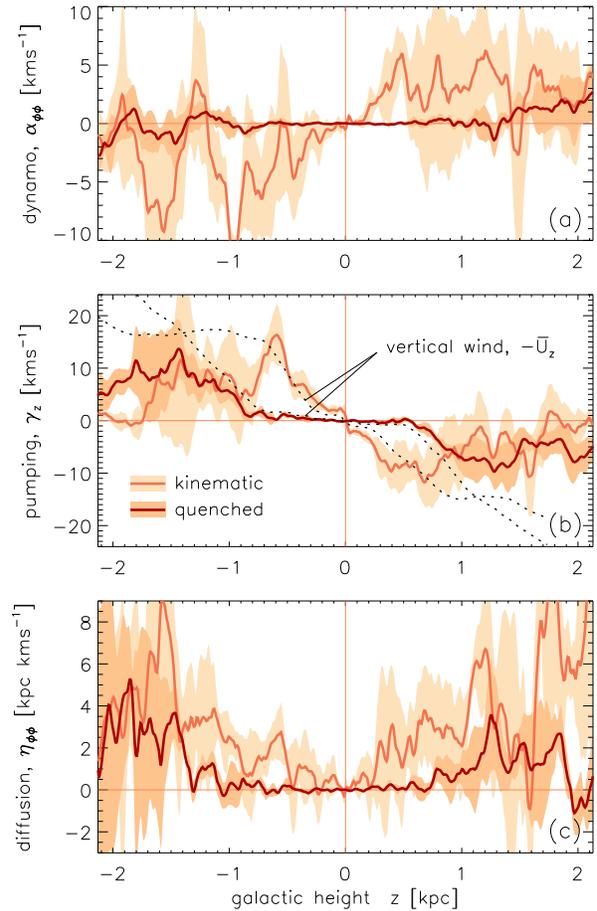}
  \caption{Time-averaged vertical profiles of the governing mean-field
    coefficients in the kinematic regime (light colours,
    $t\in[0.1,0.3]\,\Gyr$), and in the fully quenched phase (dark
    colours, $t\in[2.3,2.5]\,\Gyr$). Panels show, \textbf{(a)} the
    helical dynamo effect, $\alpha_{\phi\phi}(z)$, \textbf{(b)} the
    diamagnetic turbulent pumping, $\gamma_z(z) \equiv
    0.5\,(\alpha_{\phi R}-\alpha_{R\phi})$, and \textbf{(c)} the
    turbulent diffusivity, $\eta_{\phi\phi}(z)$. We also overlay the
    mean vertical flow (dotted), mirrored for the sake of easier
    comparison. Note that, in both regimes, $\gamma_z$ closely matches
    $-\bar{U}_z$ in the dynamo-active region, rendering saturation via
    the wind mechanism unlikely. This is in-line with the loss of
    pitch angle seen in Fig.~\ref{fig:time_bfield}.}
  \label{fig:tf_coeff}
\end{figure}

We adopt the usual mean-field closure
\citep[cf.][]{2005AN....326..787B} applicable to the local box
geometry, and assume that the turbulent electromotive force
$\EMF(z,t)$ can be approximated by vertical profiles of the mean
horizontal magnetic field $\mB=(\bar{B}_R, \bar{B}_\phi)^{\rm T}$, and
its vertical gradients, i.e.,
\begin{equation}
  \EMF_i = \alpha_{ij}\,\bar{B}_j 
         - \eta_{ij}\,\varepsilon_{jkl}\,\partial_k \bar{B}_l\,,
  \qquad i,j \in \left\{R,\phi\right\}, k=z\,.
  \label{eq:closure}
\end{equation}
We here focus on the relevant components of the tensors $\alpha_{ij}$
and $\eta_{ij}$, namely the dominant contribution
$\alpha_{\phi\phi}(z)$ to the $\alpha\Omega$~dynamo, the vertical
diamagnetic effect $\gamma_z(z) \equiv (\alpha_{\phi R} -
\alpha_{R\phi})/2$, and the turbulent diffusivity
$\eta_{\phi\phi}$. These quantities are plotted in the three panels of
Fig.~\ref{fig:tf_coeff}, respectively.

For each of the three effects, we show profiles at an early stage
(light colours), representing the (magnetically unquenched)
\emph{kinematic} regime, and at a late time, representing the fully
\emph{quenched} regime. We observe that all three profiles are
suppressed in the central region, corresponding to the belt of strong
fields seen in Fig.~\ref{fig:spctm}. We furthermore find the mean flow
$\bar{U}_z$ to be shut-off by the emerging magnetic field, explaining
the absence of quenching via the wind, and hence the low pitch
angle. This is somewhat unexpected and will require careful further
inspection -- especially since the situation appears to change at
higher supernova rate.

\begin{figure}
  \center\includegraphics[height=0.55\columnwidth]{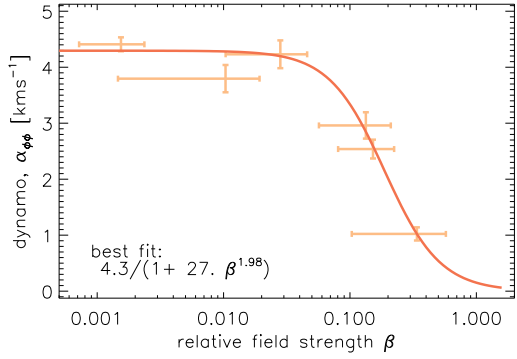}
  \caption{Amplitude of the $\alpha$ effect as a function of relative
    field strength $\beta\equiv \rms{\bar{B}/B_{\rm eq}}$.  Note that
    quenching occurs significantly below equipartition.}
  \label{fig:quen_al}
\end{figure}

The central aim of the current paper is to extract quenching
functions. We here restrict ourselves to run ``Q4'' as it presents us
with the widest range of magnetic field strengths. Comparison with
other runs, however, indicates that the derived relations should be
fairly universal. To disentangle the vertical variation from the
magnetic field dependence, we make a model assumption on the shape of
the profiles (see Sect.~\ref{sec:methods} for details). For
$\alpha_{\phi\phi}(z)$ we measure the profile slope within $|z|\le
1.33\kpc$; the resulting amplitude is plotted in
Fig.~\ref{fig:quen_al} as a function of the relative field
strength. Strictly speaking, the ordinate should have additional units
$\kpc^{-1}$, but we drop these for the sake of simplicity (and
implicitly interpret the values as integral value at $1\kpc$). With
$p_\alpha=1.98\pm 0.3$, the fitted algebraic quenching function,
cf. equation~(\ref{eq:quen}), is found in close agreement with the
commonly assumed quadratic \citep[e.g.][]{2000PhRvE..61.5202R}
dependence on $\beta$; but note that the classic result of
\citet{1972JFM....53..385M} and \citet{1974AN....295..275R} suggests a
steeper scaling with $p_\alpha=3$ -- notably for fields
\emph{exceeding} equipartition, however. In the limit of low $\Rm$,
\citet{2007MNRAS.376.1238S} have found $p=2$ for stochastic forcing,
and $p=3$ for steady forcing, which might provide an explanation. We,
moreover, find a rather large coefficient\footnote{As a potential
  caveat, we want to point out that (as speculated upon earlier) the
  $p$ and $q$ parameters are found to be correlated at the
  90\%~level.} $q_\alpha=27 \pm 14$. This is in fact cause for some
concern as one would naively expect $q_\alpha\sim 10$, based on the
scale separation $L_0/l_0\simeq 1\kpc/0.1\kpc$, as can be derived in
the framework of dynamical quenching (see sect.~9.3.1 in BS05). A
value significantly larger than this would be indicative of
``catastrophic'' quenching, for which one can estimate $q_\alpha\sim
\Rm$. In previous work, we adopted a definition of $\Rm$ based on the
global box scale, arriving at a (seemingly large) value of $\Rm'\equiv
L^2\Omega/\eta \simeq 10,000$. In the current context, it is however
more realistic to base $\Rm$ on the outer scale of the turbulence,
i.e. $k_{\rm f}\simeq 2\pi/l_0$, and a typical rms velocity,
e.g. $\urms\simeq 25$--$50\kms$. With this new definition, we obtain a
value of $\Rm\equiv \urms (k_{\rm f}\,\eta)^{-1} \simeq 75$--$125$
which is still comfortably larger than the factor appearing in the
quenching expression. Another commonly used definition of $\Rm$ in
this context is $\Rm_{\rm T}\equiv \eta_{\rm T}/\eta$. With $\eta_{\rm
  T}\simeq 0.1$ -- $2.0\kpc\kms$ in the region of interest, this makes
for a poor indicator as we obtain $\Rm_{\rm T} \simeq 15$--$300$, with
uncertain conclusions. Nevertheless, the presence of catastrophic
quenching cannot be safely ruled out here. Following the notion that
alleviating helicity constraints requires a flux of helicity out of
the dynamo-active region \citep{2002A&A...387..453K}, its occurrence
would, however, be consistent with the suppression of the wind profile
(see dotted line in Fig.~\ref{fig:tf_coeff}) at late times. Since the
wind may be the deciding factor it becomes mandatory to better
understand its exact physical origin.

\begin{figure}
  \center\includegraphics[height=0.55\columnwidth]{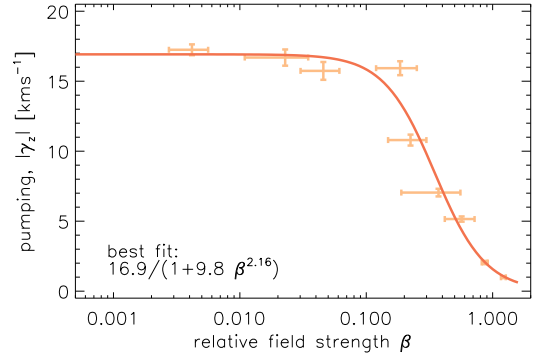}
  \caption{Same as Fig.~\ref{fig:quen_al}, but for the amplitude
    $|\gamma_z|$ of the diamagnetic turbulent pumping term. Compared
    to $\alpha_{\phi\phi}$, the pumping appears to be quenched at
    somewhat higher relative field strength.}
  \label{fig:quen_gm}
\end{figure}

For the diamagnetic term $\gamma_z(z)$, which is more centrally
confined, we infer slopes within $|z|\le 0.5\kpc$. Results are plotted
in Fig.~\ref{fig:quen_gm}, and with $q_\gamma=9.8\pm 1.6$, and
$p_\gamma=2.16\pm 0.25$, they again agree with the assumption of a
quadratic dependence $\sim\beta^2$. Like for $\alpha$, we are at odds
with the theoretical expectations of \citet{1992A&A...260..494K}, who
predict the same $\beta^3$ scaling for $\gamma_z$ -- but recall the
result by \citet{2007MNRAS.376.1238S}, mentioned above, which might
resolve this. Because diamagnetic pumping is footed on inhomogeneity
of the turbulence, but \emph{not} on lack of mirror symmetry, it is
not affected by issues of helicity conservation. This is consistent
with a low value of $q_\gamma$. We moreover performed a corresponding
fit for the mean flow $\bar{U}_z(\beta)$ and obtain comparable
parameters. This is expected from the close correspondence between the
two quantities seen in Fig.~\ref{fig:tf_coeff}.

A potential buoyant contribution to the pumping
\citep{1992A&A...260..494K} appears unlikely in our case as the Galaxy
(unlike the solar convection zone) is only moderately stratified in
density. Similarly, turbulent transport arising from small-scale
magnetic fluctuations \citep{2003GApFD..97..249R} may only become
important\footnote{Whether this is expected to counteract the
  kinematic effect is, however, dependent on rotation (cf. sect.~10.3
  in BS05).} at higher $\Rm$, due to small-scale dynamo
action. Because we find no direct evidence for the latter, we
attribute the observed attenuation to the quenching of the kinematic
effect $\gamma_z \sim -1/2 \nabla\,\eta_{\rm T}$.

\begin{figure}
  \center\includegraphics[height=0.55\columnwidth]{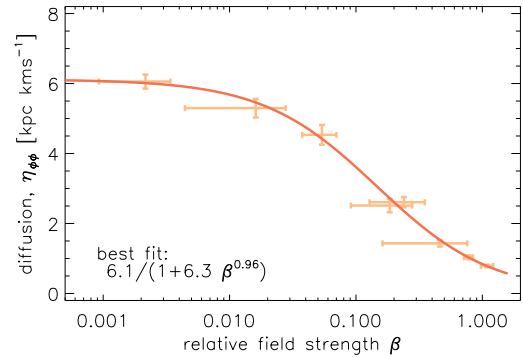}
  \caption{Same as Fig.~\ref{fig:quen_al}, but for the turbulent
    diffusivity $\eta_{\phi\phi}(\beta)$.}
  \label{fig:quen_et}
\end{figure}

To conclude the results section, we present, in
Fig.~\ref{fig:quen_et}, the quenching function for the turbulent
diffusivity $\eta_{\rm T}(\beta)$. We note that due to background
shear, the $\eta$~tensor becomes anisotropic. We however cannot
address this issue with the current data. For practical purposes, we
restrict our analysis to $\eta_{\phi\phi}$.  Unlike for the two
preceding quantities related to the $\alpha$~tensor, we now find a
scaling exponent closer to one, i.e. $p_\eta=0.96\pm 0.06$. This is in
good agreement with $p_\eta=1$, as predicted by
\citet{2000PhRvE..61.5202R}, and coincides with the scaling
suggested\footnote{Note that this value is not derived from a
  quenching function but emerges when modelling an
  $\alpha\Omega$~dynamo and comparing it to direct simulations.}  by
\citet{2002ApJ...579..359B}. Our result also agrees with the effective
diffusivity (in the one-dimensional case) quoted in the line after
eqn.~(46) of \citet{1994AN....315..157K}. \citet{2003A&A...411..321Y}
have derived diffusivity quenching from simulations of forced
turbulence. Their results appear consistent with an assumed
$p_\eta=2$, based on which they state a best-fit value of $q_\eta=8$.

% ------------------------------------------------------------------------------

\vspace{-1ex}
\section{Discussion \& Conclusions}
\label{sec:discussion}

We have provided magnetic quenching functions for the attenuation of
turbulent transport coefficients in supersonic interstellar
turbulence. Our results are based on realistic simulations of the
stratified ISM -- with the exceptions of neglecting a potential
contribution from cosmic rays
\citep[CRs,][]{2004ApJ...605L..33H,2006AN....327..469H}. Nevertheless,
the obtained quenching laws represent the first quantitative result
derived from such a complex scenario and will be highly useful in the
global modelling of the Galactic dynamo.

Generally, the obtained results appear consistent with the picture of
purely \emph{kinematic} effects, i.e., those arising out of
correlations in the fluctuations of $\urms$. These correlations are
then attenuated by the presence of a strong mean field. A key question
to address in future studies is, in how far correlations in background
\emph{magnetic} fluctuations (stemming, e.g., from a small-scale
dynamo) will be of importance. Such contributions have usually been
ignored since they are hard to capture. On the other hand, recent
studies have found that the small-scale dynamo is very hard to excite
in the multi-phase ISM, which in turn challenges its importance.

Independent of these, a potential influence of additional constraints
(arising from magnetic helicity conservation at high magnetic Reynolds
numbers) cannot safely be excluded at the current stage. Given the
considerable uncertainty in determining the related $q$ parameter, we
conclude that settling this issue will require more detailed
investigation of the scaling of our result with $\Rm$, demanding for
potentially very expensive computations. It is interesting to note
that, in comparable simulations, \citet{2012arXiv1205.3502D} found the
influence of a wind enhanced at higher $\Rm$, pointing to a possible
solution. As far as our own simulations are concerned, we will need a
thorough understanding of the conditions under which the vertical wind
is attenuated as it may play a crucial role in alleviating helicity
constraints. Because the inclusion of a CR component may potentially
resolve this deficiency, we are nevertheless confident that the
paradigm of a turbulent mean-field dynamo will prevail in explaining
large-scale Galactic magnetic fields.

% ------------------------------------------------------------------------------

\vspace{-2ex}
\section*{Acknowledgements}
We are grateful to Axel Brandenburg for giving insightful comments on
an earlier draft of this manuscript. We also thank G\"unther R\"udiger
for discussing his analytical findings, and the anonymous referee for
providing useful comments. This work used the \NIRVANA code version
3.3, developed by Udo Ziegler at the Leibniz-Institut f{\"u}r
Astrophysik Potsdam (AIP). Computations were performed on the
\texttt{babel} compute cluster at the AIP. This project is part of DFG
research unit 1254.

\appendix

% ------------------------------------------------------------------------------

\bsp

\label{lastpage}

\end{document}